*Short Research Article*

# Velvet Worms (Onychophora) in Folklore and Art: Geographic Pattern, Types of Cultural Reference and Public Perception

Julián Monge-Nájera[1]* and Bernal Morera-Brenes[2]

[1]Laboratorio de Ecología Urbana, RECAS, Vicerrectoría de Investigación UNED, 2050 San José, Costa Rica.
[2]Laboratorio de Genética Evolutiva, Escuela de Ciencias Biológicas, Universidad Nacional, Heredia, Costa Rica.
*Corresponding author: E-mail: julianmonge@gmail.com

**ABSTRACT**

**Aims:** To document and preserve folkloric beliefs and art inspired by velvet worms (Onychophora), rare invertebrates that are considered "living fossils", have full placental organs and capture prey with a rough "net" built in a fraction of a second.

**Study Design:** This study is a combination of field interviews, online surveys and automatic database search.

**Methods:** We asked open-ended questions to farmers who know the worms, consulted experts and searched the Internet to document folkloric and artistic instances using all the names that these animals receive in English, Spanish and Portu-guese (languages of the countries where they occur) as well as other languages, and automatic image search, in Web of Science, CrossRef, Google Scholar and Google.

**Results:** We found more than 80 cases of direct references to velvet worms in folklore and art, mostly from the USA, Australia, Brazil, Costa Rica and New Zealand. Per capita the countries with more cases are New Zealand (24 cases per million inhabitants), Costa Rica (16 cases per million inhabitants) and Australia (3 cases per million inhabitants). The most frequent expressions are cartoons, followed by tourism agencies using velvet worms in their ads, products with velvet worm representations, folkloric beliefs, and music bands or songs named after them. In almost all cases the animals are seen in a favorable light, inspiring folklore and art that highlight their extraordinary nature.

**Conclusion:** The unique prey capture mechanism of velvet worms seems to have inspired an unexpected number of artistic and folkloric expressions, preserved for the future in the present article, which starts a totally new line of research: the effect of "living fossils" on human culture.

## 1. INTRODUCTION

Velvet worms (phylum Onychophora) are terrestrial invertebrates with a superficial resemblance to caterpillars and very low population densities in most habitats, so they can be labeled as "rare". Living species occur in tropical and southern-temperate ecosystems in isolated regions of the planet, where they usually inhabit dark, moist microhabitats, mainly in soil and forest litter, rotting logs and bromeliads. Their name originates in peripatoí, Greek for itinerant, associated with Aristotle and his peripatetic school, and refers to their particular gait based on short legs that have no joints [1].

Because of their rarity, few people –including professional biologists- have had the experience of observing them in their natural habitat (it is surprising that one people, the Chuave of Papua New Guinea included them in their diet, according to Meyer-Rochow [2]. Even though they never become crop pests or transmit human disease, these animals are of scientific importance because they are an ancient group (Cambrian) and have characteristics in common with annelids and arthropods [3]. Tropical species give birth to live young [4] and appear to be the only invertebrates in the world with a full placental organ [5]. On the other hand, they have been considered "living fossils" because their anatomy has changed very little over time [6]; the term was coined and explained by Darwin himself in p. 107 of On the Origin of Species (1859 edition): ".. some ... anomalous forms may almost be called living fossils; they have endured to the present day, from having inhabited a confined area, and from having thus been exposed to less severe competition".

Despite their frail appearance, they are carnivores and hunt small invertebrates such as worms, mollusks, isopods, spiders, crickets and termites. The mechanism by which they capture prey is unique in the animal kingdom [3,7]. A velvet worm walks on the forest litter in search of food at night or in very moist days; when it perceives a potential prey -sometimes corroborating with a soft touch with the smell sensors in its antennae- the worm ejects two powerful jets of liquid glue through specialized organs ("turrets") on the sides of the head. Concha et al. [8] recently described the physical details of this mechanism: thanks to a fast oscillatory movement of the glue turrets, the streams cross in mid-air, weaving a disordered self-assembled net which immobilizes its target. The liquid hardens quickly in contact with air and when the prey is entangled and secured with adhesive, the worms use two pairs of jaws to open the exoskeleton and consume the contents with external digestion. The squirting mechanism is also used for defense from predators, sometimes reaching a distance of 70 centimeters [9].

These worms have been known to science for two centuries, but there are no studies of cultural expressions referring to them, with the exception of a paragraph in a newsletter by Monge-Nájera [10]. Velvet worms are not well known to the general public and in Spain there was a protest when a question about them appeared in an exam (elpais.com: http://goo.gl/dC3r1b). They lack names in most cultures. In Costa Rica the few people who are familiar with them consider velvet worms slugs (babosa in Spanish); they might add that they are special because, unlike other "slugs", they have legs: babosas con patas [10]. In the north of Brazil they are called Embuá de chifre

Monge-Nájera and Morera-Brenes; BJESBS, 10(3): 1-9, 2015; Article no.BJESBS.18945

(millipeds with long antennae: Cristiano Sampaio, personal communication 2014). In Belize the Maya Mopán call them Ko-mes and in New Zealand the Māori name is Ngaokeoke, from ngaoki, "to crawl"(www.teara.govt.nz/en/peripatus).

There was an "explosion" of cultural references to velvet worms after 2005, most of them online. Here we document velvet worms in folklore and other aspects of popular culture. Our motivation is to preserve for the future local folklore and Internet art about these worms. The stories we heard from old farmers will be lost because their way of life is disappearing. Internet art is published in volatile forums that will disappear when older posts are erased from servers. We cover expressions from poems and music bands to mugs and t-shirts, and try to understand why velvet worms are seen in a positive light despite belonging to the generally unpopular category of "worms". Furthermore, we examine the role that television played in their entrance to general culture and how their prey capture mechanism appears to have affected their perception by the general public. There is absolutely no previous literature on how velvet worms influence human cultural productions such as art and folklore. Thus the main significance of this article is that it starts a new line of research: The effect of "living fossils on human culture.

## 2. METHODS

For the search we used the terms velvet worm, peripato, peripatus, onychophoran and Onychophora (i.e., their names in languages of the countries where they occur: English, Spanish, Portuguese), as well as equivalents in Chinese, Japanese and Russian. To avoid language limitations, we also used automatic image search. For this we used Web of Science, CrossRef, Google Scholar and the more comprehensive search motor google.com (i.e., American version of Google, October 2014 through April 2015). We checked all the pages of results until they no longer had pertinent results. We also asked open-ended questions to Costa Rican farmers who know the animals and e-mailed all people who had produced poems, paintings, drinks, recipes, crafts, videos, songs, videogame characters or any other cultural expressions mentioning velvet worms, or had names related to them (total N=90 people). For this we used the contact information in their publications (and searched for email address using their names in the few cases when there was no contact information). We asked each for additional information on their specific cultural expression, how they had learned about velvet worms, and why they had chosen to produce cultural expressions inspired in these invertebrates. We did not insist in the case of people who failed to answer because our previous experience is that insisting does not produce useful results. In our message we explained that we were doing this study to identify and classify cultural expressions about velvet worms and that we would publish the results. For those who answered and sent quotable information, we obtained written permission to publish the quote.

We also sent an online survey to 14 authors who published scientific articles about velvet worms from 2002 to the present and posted a call for myths, anecdotes, etc. in the Biodiversity Group of Linkedin.com and in answers.yahoo.com. Formal articles in magazines and newspapers are a different subject and are not part of this study. For folklore about velvet worms we searched Thompson's Index [11] with the keywords peripatus, onychophoran, velvet worm, millipede, slug, skin, glue, adhesive, squirt and wart.

To test if the increase in references was an artifact of Internet growth, we organized the data in a 3x2 table corresponding to 1, 9 and 46 products against 210, 678 and 1112 million web users in the early, mid and late parts of the period we studied, and applied a Chi-Square test. This matched data characteristics and test requirements better than a correlation and was based on the null hypotheses that product growth paralleled Internet growth. Internet growth data are from the International Telecommunications Union (http://goo.gl/3BWTcc, consulted March 26, 2015).

A full Excel file with all cases, including country, date, URL and images is freely available online under DOI: 10.13140/RG.2.1.4720.8484.

## 3. RESULTS

We found 85 cases of direct velvet worm references in human cultural expressions (Fig. 1).

By total number, most are from a country where there are no living velvet worms, the USA. A second group of countries, all with living species, have around ten cases each (Australia, Brazil, Costa Rica and New Zealand). A third group, with less cultural expressions, includes countries with living species such as Chile, South Africa, Singapore and New Guinea, and without them including Canada, some European countries and Japan.

Per capita (number of cases/million inhabitants) the countries with more cases are New Zealand (24 cases per million inhabitants), Costa Rica (16 cases per million inhabitants) and Australia (3 cases million inhabitants).

The most frequent expressions are cartoons, followed by tourism agencies using velvet worms in their ads, products with velvet worm representations, folkloric beliefs, and music bands or songs named after them (Appendix 1).

In almost all cases the animals are seen in a favorable light (only six present them in a negative way), and they inspired artistic expressions and commercial products that highlight their extraordinary nature.

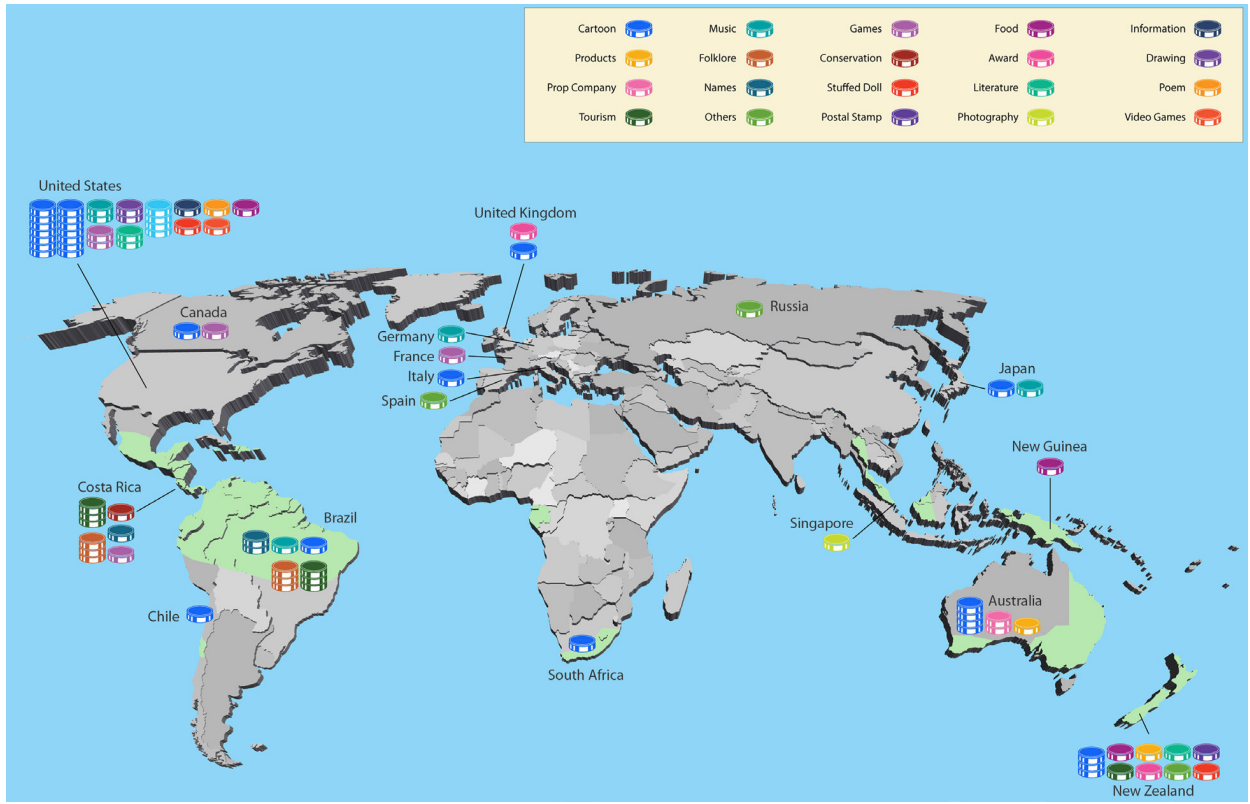

**Fig. 1.** Cases of direct references to velvet worms in human cultural expressions by type of expression and country. Each disk represents one expression. Light green in the map: Regions that currently have velvet worm species.

### 3.1. Folkloric Beliefs and Uses

As expected from rare and secretive animals, cultural expressions that refer to them are rare and quite different from place to place. They are believed to cure warts and make snakes shed their skins. They are also feared for their secretion, lumped with slugs and used as bait by fishermen. At a more sophisticated level, they can be metaphors for small, local and unique enterprises and used as example of the need to protect non-charismatic species.

In Cascajal de Coronado (San José province, Costa Rica), one farmer reported that their adhesive secretion "eliminates warts" and another explained that the glue of these worms can "make snakes lose their skin".

A local tourist guide from Bahía Ballena, Costa Rica, told us that a velvet worm squirted glue on the arm of a child, leaving a brown stain that could not be easily washed away with soap and water. He feared the liquid could be harmful (Rony Duarte, personal communication, 2014).

In rural areas of Brazil they are known to some people, who have ideas about their place in nature, similarities to slugs and usefulness as bait:

"In Brazil, some species occur next to small villages (e.g., E. diadenoproctus, E. paurognostus and E. adenocryptus) and people usually know them very well (they call them legged slugs). Once, a local told me that he always finds them among the straw used as fertilizer for coffee plantations. He also told me that he collects them and throws them back into the forest because they do not cause any damage and their place is in the forest.

Another local told me that he sometimes finds them walking at home and they usually leave a slimy track behind. I first thought he was talking about a common slug. Then, I tried to explain that he might have mixed the animals, for example, confusing a velvet worm with a common slug. He said angrily: «No! Wait a second». After 5 minutes he returned with an Onychophoran on his hand. I was completely surprised, even though I still cannot explain the "slimy track" that they "leave behind" described by him.

In the Ecological Station of Tripuí (Ouro Preto, Minas Gerais, Brazil) the type locality of E. acacioi, I was told that local fishermen used to use these animals for fishing. This makes sense, as this species lives deep in the soil and I believe that the locals ended up finding them while looking for earthworms for fishing. Nowadays, it might not happen any longer, as everybody around knows the species and is aware of its threatened status».

*Ivo de Sena Oliveira (2014, personal communication).*

The case of a velvet worm species with political and economic repercussions is unique. In New Zealand, a highway course was changed to protect Peripatoides novaezealandiae. This inspired the owner of a small company (Velvet Worm Beers) to use the worm as a metaphor of how he wanted his company to be perceived:

> "I learned about velvet worms because I was living in a suburb of Dunedin, New Zealand called Caversham, where there is a ... unique species essentially only found in a few square kilometer area around my house. I had ... even found one inside my own house living in a pot plant! Then their profile got raised in local media as there was a widening of the highway happening, which cuts through their habitat. The outcome in the end was successful ... land donated by the Land Transport NZ has become an official reserve ... adjacent to a private property which the owner has used as an informal velvet worm reserve for some years. He builds habitats for them out of piles of bricks or logs, and has been planting native trees and shrubs to help their habitat. Another great outcome of this process was that it inspired the local branch of the Department of Conservation to make a resource booklet about Peripatus, which not only details them from a zoological/ ecological perspective, but also talks about the process that went on with regards to the roading project and how stakeholders etc. were involved ... So basically when I started my own microbrewery, having had an interest in all things biology for a long time, I thought that the Velvet Worm was a great way to express my brewery, which is small, unique and local, just like our local species of velvet worm".

*Bart Acres (personal communication, 2014)*

Finally, these worms have been used to send an anti-discrimination message. DeviantArt member Soyrwoo, self described as a 20 year old woman from the USA, painted a character named Victoria, the pink velvet worm: "Rather than using the 'cute and cuddly' critters of the world as mascots, NaLoA uses 'ugly' species to encourage the conservation of all habitats and endangered species. In order to do this, the six 'ugly' species are 'cutified' through cartooning and slapped onto T-shirts and mugs, with the royalty going to the World Wildlife Fund" (deviantart.com: http://goo.gl/dC3r1b).

## 3.2 Possible Relationships

We did not find online cultural expressions before 2005, the year that velvet worm hunting behavior was shown in the BBC documentary Life in the Undergrowth. The pattern differs significantly from the growth of Internet users (Chi-square 22.45; 5 degrees of freedom; p=0.0004).

Only five people answered our question about why they had produced some cultural expression about velvet worms. One said the she had been inspired by a book and four that they had been inspired by this particular documentary. The poem Perspective as learned from the velvet worm of New Zealand originated in the idea that some velvet worms are brightly colored but lack the visual capacity to see each other's colors, taken from a book (Christina Olson, personal communication, 2014). A forum poster to DriveThruRPG.com proposed a velvet worm character for a videogame because, despite its "poor defense" it had "animal intelligence; superb stealth and hide [sic] abilities; great strength, sticky slime attack (20' range, 5' radius) and good bite attack with fair acidic saliva", explaining "This creature idea was stolen directly from a BBC documentary on insects I watched recently, Life in the Undergrowth". (http://rpgdump.blogspot.com/2009/02/velvet-worms.html).

A design company in Australia that produces masks, fake weapons and similar products is named Velvet Worm Designs because "we couldn't think of a cool name that wasn't already taken and we watched a lot of Attenborough; one day it just clicked" (Jon Toll, personal communication, 2014).

The name of the Japanese death metal band Velvet Worm was chosen after watching the same documentary "because the word velvet has a nice ring and because of the violent way in which the animal captures its prey" (Miss Junk –artistic name– singer and guitar player; personal communication through T. Taenaka and K. Nishida, 2014).

Science blogger Bethany Brookshire noted that the video for Katy Perry´s 2010 song California Gurls ft. Snoop Dogg (youtube.com/watch?v=F57P9C4SAW4) shows shooting turrets that resemble those of velvet worms "as they appear on the Life in the Undergrowth video"

(http://scicurious.scientopia.org/2011/02/25/friday-weird-science-killin-prey-with-my-super-scary-glue-gun/).

## 4. DISCUSSION

Parts of this study, such as folkloric beliefs about velvet worm secretion curing warts and forcing snakes to molt can be situated in the tradition of narrative inquiry, i.e., the study of human aspects of culture that cannot be reduced to dry facts and numbers [12]. Thompson's Index [11] has no cases of velvet worms in folklore or about their alleged capacity of curing warts. Warts are related to viruses and their effects easily appear and disappear; this has resulted in a large number of beliefs about what can cure them [13] and now velvet worm adhesive must be added to the list of remedies. Snakes shed their skins periodically: possibly someone put velvet worms and snakes together in a bag, saw the snake discard its old skin, and thought one thing caused the other [10].

The story about the hard to wash, colored and possibly harmful stain reflects a common fear about animals that might be dangerous [14], while the association with slugs reflects a real resemblance that also misled early naturalists, who thought that velvet worms (then recently discovered in the Caribbean) were a new type of mollusk or at least an animal very closely related to mollusks [15]. In this particular belief, tropical farmers are not different from the learned naturalists of the old days: the unique nature of velvet worms required internal anatomical studies to become apparent and to lead to the conclusion that they were so unique that they deserved a whole phylum of their own [16].

Velvet worms capture prey with a non-lethal net that can explain why a worm is seen in a positive light. The police and the military have long wanted such an efficient non-lethal weapon that currently only exists in fiction (e.g. the Star Wars novels Thrawn Trilogy [17,18]; and even though there is not a single spider species that produces an instantaneous web [19] superhero Spider-Man [20] produces a velvet worm-like net to capture enemies and to help friends.

In 2005 a slow motion video of Peripatus solorzanoi capturing prey with its non-lethal net produced by the BBC was broadcasted by two television channels that are seen by millions of people the world over, Animal Planet and Discovery Channel and by the BBC itself [21]. The replies from the people whom we contacted and the dates agree with the hypothesis that many of the cultural expressions we found were inspired by this documentary. We did not find any information suggesting the there was another reason but of course that is also a possibility; we only state that the data available to us are in agreement with that hypothetical explanation. The non-lethal but spectacular prey capture mechanism [22] could explain why the animals were mostly perceived in a positive way. Again this is just a hypothesis that fits the data we were able to obtain and other possibilities might be found by future researchers. It is general knowledge that worms are seen with disgust by most people, but even so a species can have both positive and negative associations in human culture, according to a review of animal symbolism and human attitudes published recently by Benavides [23].

The predominance of cultural expressions from the USA is not surprising: the USA often comes first in any study of cultural influence because of its large population and powerful cultural industry, as found by previous researchers [24]. The many cases from New Zealand and Costa Rica can be explained by the presence of these animals, which occur in those countries, in the local media, particularly in a conservation debate in New Zealand widely covered by the press [25,26], and the fact that Costa Rica is a country with a strong conservationist culture where television and newspapers have repeatedly covered velvet worms [27-31].

The key contribution of our study is that it provides the first-ever analysis of how the phylum Onychophora is present in human culture worldwide, with a sample of more than 80 documented cases and consideration of oral accounts of folkloric beliefs about these animals. It also preserves information that otherwise would be lost for future researchers of zoologically inspired art and folklore.

## 5. CONCLUSION

Folklore about velvet worms is among the least known cases of zoologically inspired folklore. Despite being unknown to much of the general public before the broadcast of Life in the Undergrowth, the combination of an unique prey capture mechanism (shown in spectacular slow motion detail in the documentary), and the wide reach of international television, seems to have inspired people around the world to produce an unexpected number of cultural expressions based on what arguably might be the most extraordinary of all "living fossils". The study of invertebrates in human culture is particularly poor and even subject to prejudice, as reported by Stix, Stix and Abbot, who found strong opposition while attempting a similar study about mollusks in art. Scientists and artists, limited to the narrow views of their fields, failed to see the value of what is now a respected classic, The Shell: Five Hundred Million Years of Inspired Design [32]. We hope this article will reach a similar goal for worms by documenting these cultural expressions before they disappear and that it serves as a baseline for future researchers of this novel aspect of cultural anthropology.

## ACKNOWLEDGMENTS

We thank the following for information cited in this article: Bart Acres, Tomomi Taenaka, Junk (singer, Velvet Worm band), Kenji Nishida, Christina Olson, Cristiano Sampaio, Ivo Sena, Patricia Valverde, Paul Whitington, Rony Duarte, Jon Toll, and Andréa Peripato (who graciously explained that her family name is of Italian origin and probably unrelated to the Onychophora). We also thank Priscilla Carbonell and Alonso Prendas for assistance in producing the figures and Zaidett Barrientos for the statistical analysis, as well as Pelayo Benavides (Pontificia Universidad Católica de Chile), Lawrence Millman and two anonymous reviewers for very useful suggestions to improve the manuscript. B.M.-B. was supported by Universidad Nacional de Costa Rica, Project 0095-14.

## COMPETING INTERESTS

Authors declare no competing interests.

# APPENDIX 1

Excel spreadsheet with all the data.

Dataset

Animals-Monge-Nájera rare living fossils-DataBase
Julián Monge-Nájera, Bernal Morera-Brenes.
DOI: 10.13140/RG.2.1.4720.8484
https://www.researchgate.net/publication/274393253_animals-Monge_rare_living_fossils-DataBase

Peer-review history:
The peer review history for this paper can be accessed here:
http://sciencedomain.org/review-history/10062